\documentclass[fleqn,usenatbib]{mnras}

\usepackage{newtxtext,newtxmath}
\usepackage[T1]{fontenc}
\DeclareRobustCommand{\VAN}[3]{#2}
\let\VANthebibliography\thebibliography
\def\thebibliography{\DeclareRobustCommand{\VAN}[3]{##3}\VANthebibliography}
\usepackage{graphicx}	
\usepackage{amsmath}	
\usepackage[linesnumbered, noend]{algorithm2e}
\usepackage{algpseudocode}
\usepackage{enumitem}
\setlist[enumerate,1]{label=\arabic*., left=4pt, labelsep=2pt}

\graphicspath{{./}{figures/}}

\title{
Constructing Merger Trees of Density Peaks Using\\ Phase-Space Watershed Segmentation Algorithm
}
\author[Geda and Teyssier]{
Robel Geda,$^{1}$\thanks{E-mail: robel@princeton.edu}
Romain Teyssier,$^{1}$
\\
$^{1}$Department of Astrophysical Sciences, Princeton University, Princeton, NJ 08540, USA\\
}
\date{Accepted 22 December 2024}
\pubyear{\the\year{}}

\begin{document}
\label{firstpage}
\pagerange{\pageref{firstpage}--\pageref{lastpage}}
\maketitle

\begin{abstract}
Structure identification in cosmological simulations plays an important role in analyzing simulation outputs. The definition of these structures directly impacts the inferred properties derived from these simulations. This paper proposes a more straightforward definition and model of structure by focusing on density peaks rather than halos and clumps. It introduces a new watershed algorithm that uses phase-space analysis to identify structures, especially in complex environments where traditional methods may struggle due to spatially overlapping structures. Additionally, a merger tree code is introduced to track density peaks across timesteps, making use of the boosted potential for identifying the most bound particles for each peak.
\end{abstract}


\section{Introduction} \label{sec:intro}

Identifying structure in cosmological simulations represents a key step in analyzing simulation outputs. How these structures are defined and identified influences the physical properties that are derived from the simulations. These properties are subsequently used for comparison with observational data \citep{2011MNRAS.415.2293K}. Furthermore, it is important to track structures across multiple timesteps so we can track their evolution and dynamics. As such, dark matter halos are important structures to identify and track since they serve as the basic units in the hierarchy of structures that form the large-scale universe.

Dark matter halos are virialized systems that are dominated by dark matter. Most simulations use dark matter particles to model the formation and evolution of such structures. Specifically, a dark matter halo is defined as a concentrated and virialized clump of dark matter particles. Naturally, the density of these particles produces a potential well, which is often used to identify which dark matter particles are bound to the system in a process called binding checks. This process is important for determining the masses of dark matter halos but can prove cumbersome, especially when the potential landscape becomes increasingly complicated with multiple overlapping structures \citep{boosted}. Furthermore, The mass and size of a dark matter halo can vary based on the definitions and parameters used, such as the total number of bound particles or the choice between using the virial radius, the peak of the rotation curve \citep{2008MNRAS.386.2022A, 2011MNRAS.415.2293K}, splashback radius \citep{2014ApJ...789....1D, 2014JCAP...11..019A, 2015ApJ...810...36M}, or their tidal boundaries \citep{boosted}. Given the non-negligible impact of chosen parameters on halo properties, this paper advocates for a shift towards focusing the primary goal of structure finders on density peaks rather than the dark matter halos themselves.

Since density peaks often occur at the centres of halos and subhalos, we can achieve a more detailed and unbiased phase-space assessment of the density field compared to mass-centric methods. Simulations can use the locations of peaks to inform black hole seeding in simulations \citep{2005MNRAS.361..776S, 2007MNRAS.380..877S, 2014MNRAS.442.2751T, 2009MNRAS.398...53B, 2017MNRAS.470.1121T}, clustering statistics of galaxies \citep{2018PhRvD..98d3526A, 2024A&A...682A.148F}, and pinpointing of galaxy cluster centres (e.g. DENMAX clusters \cite{2003MNRAS.339..271S} or \cite{2024arXiv240509855M}). Studies have also shown that density clustering statistics and shear peak statistics provide strong cosmological constraints, improving parameter precision compared to using mass functions alone \citep{2018PhRvD..98d3526A}. Finally, once the peaks have been located, the extent and masses of the structures that surround them can be determined as an optional post-processing step. For example, using the peak locations as starting points, the parameters of dark matter halos can be measured by fitting Navarro-Frenk-White \citep[NFW hereafter]{1996ApJ...462..563N} profiles, checking the binding energy of particles using (boosted) gravitational potentials \citep{boosted}, or constructing radial profiles until they fall below a specified density threshold.

Like dark matter halos, peaks can be traced across different time steps, and a merger tree can be created for them. Since these peaks typically endure even through mergers, density peaks provide a dependable way to monitor sub-structures within halos. In conventional methods, a halo is deemed lost if it loses a significant amount of mass and is seen as having merged with a larger structure. This so-called overmerging problem \citep{1987Natur.330..451W, 1996ApJ...457..455M, 1998MNRAS.299..728T, 1999MNRAS.310...43V, 1999ApJ...516..530K, 2000ApJ...544..616G, 2011MNRAS.410.2309G} can in some part be alleviated by considering density peaks even after the halos that surround them have been largely stripped. In addition to better tracking, this would allow us to study the formation and evolution of what are widely believed to be merger remnants. Such objects include ultra compact galaxies in clusters and the Omega Centauri cluster within the Milky Way.

Many algorithms have been developed for structure identification since the advent of cosmological simulations \citep{1974ApJ...187..425P}. One such algorithm is the friend-of-friends algorithm, introduced in \cite{1985ApJ...292..371D}, which identifies structures based on the spacing between particles. Briefly, the algorithm works by assigning a given particle to a group if it is within a characteristic linking length, usually $0.2$ times the mean inter-particular distance, of another particle already assigned to that group. Building on this concept, advanced algorithms like ROCKSTAR \citep{2013ApJ...762..109B} have extended the principle into phase space, where both positions and velocities of particles are considered. This adaptation allows ROCKSTAR to identify and ``deblend" structures that may overlap in configuration space but are separable in their dynamical states. Other popular structure and sub-structure finding algorithms include, but are not limited to, the Spherical Overdensity Method \citep{1992ApJ...399..405W, 1994MNRAS.271..676L, 2022MNRAS.509..501H}, SUBFIND \citep{2001MNRAS.328..726S}, HOP \citep{1998ApJ...498..137E}, AMIGA Halo Finder \citep{2004MNRAS.351..399G, 2009ApJS..182..608K}, and Watershed Segmentation \citep{PHEW}. 

As simulations grow in complexity and become computationally intensive, running structure identification and merger tree algorithms alongside simulations has become an appealing option. This so-called on-the-fly approach is gaining popularity due to the efficient use of the same computational resources as the underlying simulation. Consequently, optimizing the merger tree code for speed and efficiency is important to minimize costly overheads. It is with this in mind that codes like Parallel Hi Erarchical Watershed (PHEW) \citep{PHEW} and its corresponding merger tree code ACACIA \citep{ACACIA} were implemented into the octree-based adaptive mesh refinement (AMR) code \texttt{RAMSES} \citep{2002A&A...385..337T}. PHEW applies a watershed algorithm to the AMR grid to identify clumps of particles. Though this approach is efficient and powerful for identifying structure, it is limited in that it only considers configuration space. 

In this paper, we introduce a new watershed algorithm that extends the capabilities of PHEW by focusing on density peaks and incorporating phase-space information. This shift allows for detecting structures in complex environments where traditional methods may fail due to spatially overlapping structures. Furthermore, we introduce a merger tree code to accompany our definition of structure. In the past, structure finders have used some form of  {\it self-potential}, meaning the potential of the clumps as if they were in isolation, for computing binding energies. Inspired by \cite{boosted}, we use the much more meaningful {\it boosted potential} to identify the most bound particles in each density peak and use them to track density peaks across timesteps. Furthermore, because we are focused on the peaks, it is unnecessary to define the boundaries of the clumps (using the tidal radius) in the structure finding step and subsequent merger tree steps.

In Section \ref{sec:Density_Peaks}, we define the structures we aim to identify and track, namely density peaks. In Section \ref{sec:density_peak_finder}, we discuss the peak finder algorithm and the accompanying merger tree code in Section \ref{sec:merger_tree}. Lastly, we demonstrate these algorithms in Section \ref{sec:Demonstration and Comparison}. 

\section{Density Peaks} 
\label{sec:Density_Peaks}

In this section, we strictly define density peaks in the context of simulations. A density peak is a localized region where the density of matter, typically dark matter, exceeds the surrounding density by a significant margin. A density peak is parameterized by seven coordinates: six phase-space coordinates and one time coordinate, which we refer to as the ``peak coordinates". While additional information, such as the potential and density at the peak,  can be linked to the peak, we consider these types of attributes to be secondary to the peak coordinates. 

In addition to the density peak coordinates, it is necessary for a peak to be associated with bound particles. The quantity of particles comprising a peak will differ based on the nature and resolution of the simulation. However, each peak should be represented by a group of the most bound particles. This requirement is helpful because it ensures that the peak is not merely a transient feature of the density landscape, and the most bound particles can be used to track the continuity of the peak across timesteps.

Density peaks arise in scenarios such as simulations of the interstellar medium and large-scale cosmological simulations. Though these peaks exhibit varying behaviours in different contexts, they are universally characterized by a peak coordinate and a group of most bound particles. This paper focuses specifically on collisionless dark matter particles at dark matter halo scales and their density peaks.


\section{Phase-Space Density Peak Finder}
\label{sec:density_peak_finder}

\begin{figure}
\includegraphics[width=8cm]{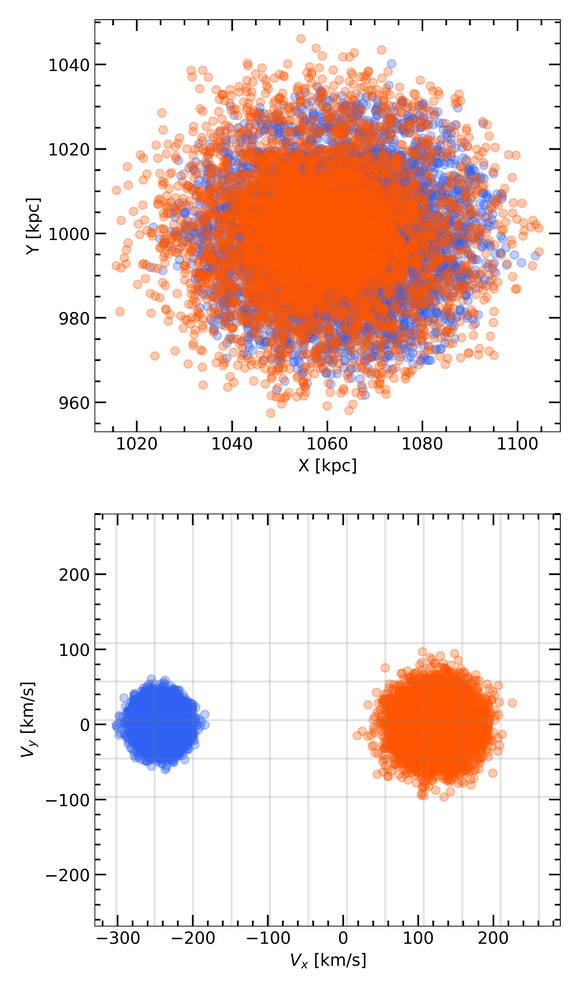} 
\centering
\caption{\label{fig:velo_deblend}
A demonstration of velocity deblending using two halos passing through each other in the x direction. Top panel: A projection of particles' x and y positions in two halos that are completely overlapping. The particles are coloured according to their halo membership. Bottom panel: the x and y velocities of the same particles. Though it is impossible to distinguish the particles' membership using their positions, there is a clear distinction between the two populations when viewed in phase space. We also show a grid in grey, defined using a fraction of the escape velocity of the total structure, that is used to bin and segment the velocities into two populations.}
\end{figure}

\subsection{PHEW Peak Finder}\label{subsec:PHEW}

We start the process of finding peaks using the PHEW structure finding algorithm introduced in \cite{PHEW_Old} and \cite{PHEW}. PHEW operates on the density field defined on the adaptive mesh and can thus be used on the gas or the dark matter particles \citep{PHEW}. The watershed algorithm first segments the computational volume into dense regions (``patches'') by following the steepest gradient, and then merges the segmented patches based on the saddle point topology of the density field \citep{PHEW}. As such, the patch of particles associated with a peak is called a ``peak patch''. PHEW is capable of automatically detecting connected regions above an adopted density threshold, as well as the entire set of substructures within. The \texttt{RAMSES} implementation of PHEW is parallelized using the MPI library and runs on the fly, which conserves memory and makes effective use of the computational resources used to run the simulation. 

To find density peak patches, PHEW first identifies all cells above a provided density threshold and marks them as ``test cells``, meaning the subset of cells that will be organized in peaks and peak patches. It then sorts the test cells by decreasing density and assigns them a patch ID. Iterating the sorted list starting from the densest cell, PHEW searches for the densest neighbouring cell for every test cell. If a test cell has a denser neighbour, the patch ID of that neighbour is assigned to the test cell, which results in the propagation of the peak cell ID to its surrounding cells. The sorted order of the cells ensures that the denser neighbour is assigned a patch ID before assessing a less dense test cell. After identifying the densest saddle points between neighbouring peak patches, PHEW eliminates peaks for which the ratio of peak density to saddle point density (``relevance threshold``) is too low by merging them with the neighbour with their densest saddle point. The typical value for the relevance threshold we adopt for cosmological simulation is 3. PHEW provides a catalogue of relevant peak patches, which we will refer to as a ``PHEW patch", and the corresponding cell and particle membership. The next step is to deal with peaks that overlap in space but are, in fact, distinct in velocity space. 

\subsection{Velocity Deblending}
\label{subsec:deblending}

To construct a reliable peak finder, we must consider density peaks overlapping in space but with enough kinetic energy to separate eventually. To achieve this, we must take the phase-space distribution of the particles into consideration. If we observe a clustered population of particles with velocities that significantly deviate from the primary population, there is likely a distinct unidentified peak within the primary peak patch. We refer to this process of isolating and separating unidentified peaks as ``velocity deblending."

We test each PHEW peak for unidentified peaks that may be present within its patch. When deblending a peak, we start by making a histogram of the velocities of the member particles. Because structures in a simulation have a diverse range of masses and velocities, we have to define our velocity bins carefully. One approach to solving this issue is to define a velocity bin size that can resolve the escape velocities for the smallest structures in the simulation. This approach, though simple to implement, becomes less feasible due to high memory requirements and significant computational costs associated with creating large histograms for massive structures. Furthermore, it's important to note that this computational inefficiency dissuades us from segmenting the simulation volume directly in phase-space (i.e. using a 6D watershed algorithm). Therefore, the velocity bin size must be defined dynamically according to the mass of each structure a peak is embedded in. We approximate the mass of the primary PHEW patch identified by PHEW, which is not the same as a halo mass, by summing the masses of all member particles. While this is by no means a perfect method of measuring the enclosed mass, it provides a sufficiently accurate approximation for estimating the escape velocity of the total structure. We use the patch mass ($M_{\rm p}$) to directly calculate the escape velocity ($v_{\rm e}$) at the position of the furthest member particle from the primary peak's centre ($R_{\rm max}$):

\begin{equation}
     v_{\rm e} = \sqrt{\frac{2 G M_{\rm p}}{R_{\rm max}}}
\end{equation}

After constructing the velocity histogram using the escape velocity (or some fraction) as our bin size, we use a watershed algorithm to identify clumps that significantly deviate from the primary bulk velocity. When multiple patches are identified, we assign particle memberships through applying a watershed segmentation on the velocity distribution. We then compute the identified peaks' central positions and bulk velocities using the particle data (i.e., mass-weighed mean). We keep track of all deblended peaks, including the primary ones, for future steps. Figure \ref{fig:velo_deblend} shows an example of two halos, coloured orange and blue, passing through each other in the x direction, but at a timestep where they are spatially indistinguishable because they overlap.  

\subsection{Artefacts}
\label{subsec:artefacts}

\begin{figure}
\includegraphics[width=8cm]{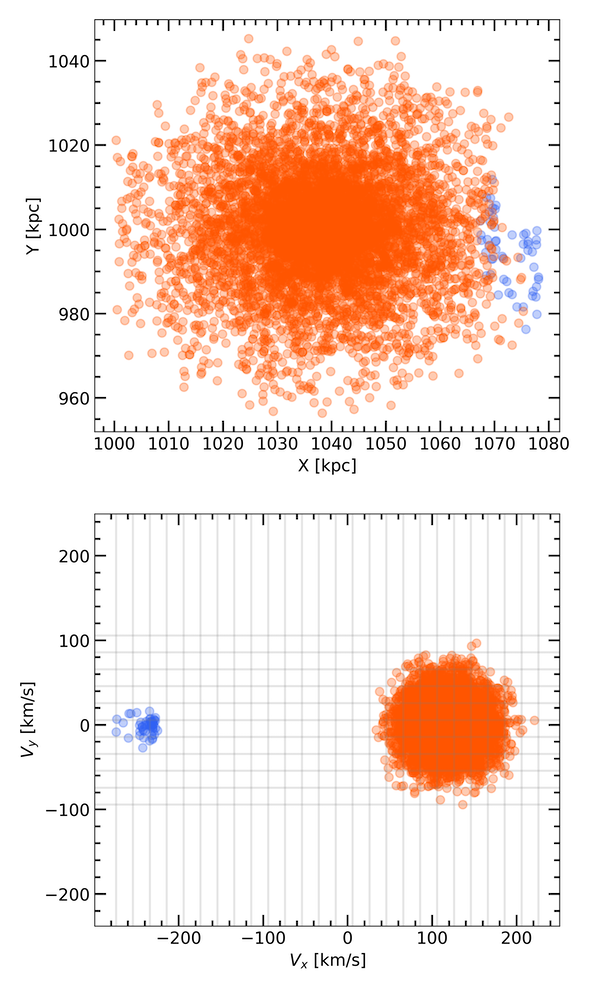} 
\centering
\caption{\label{fig:contam_velo_deblend}
Artefacts observed during velocity deblending are shown in a snapshot where two halos begin to overlap (with panels similar to Figure \ref{fig:velo_deblend}). While some particles from the blue halo are initially identified as belonging to a distinct peak, their true peak (not displayed here) lies outside the orange peak patch and is identified in the spatial watershed step. These particles are at risk of being misidentified as a new, erroneous peak if the existence of the parent peak patch is overlooked. When constructing the merger tree, we prevent such particles from erroneously being labelled as a new peak by asserting that no new peak can form within another peak's patch during the velocity deblending step.}
\end{figure}

Velocity deblending can find particle clumps within a peak patch with significantly different bulk motions. Though this indicates that there is a population of particles that may not belong to the primary peak patch, it does not necessarily mean that there is a peak within the primary associated with them. For example, consider two orange and blue structures on a collision course as shown in Figure \ref{fig:contam_velo_deblend}. Just before the blue peak enters the orange patch, some particles from blue patch will overlap with orange patch. During the deblending of orange, these contaminating particles from blue are recognized and separated as distinct peaks within orange’s patch. However, this identified structure is an artefact (or artefact) with no peak other than blue. This issue is resolved by simply asserting that no new peak can form within another peak patch by virtue of velocity deblending. In other words, a new peak can only emerge inside the patch of another peak if it is pronounced enough that it is identified spatially. The peak finder itself does not differentiate between genuine peaks and artefacts; this determination is done during the merger tree step since we need information from two different snap-shots.


\section{Constructing the Merger Tree}
\label{sec:merger_tree}

\subsection{Voter System}\label{subsec:voter_sys}

\begin{figure}
\includegraphics[width=8cm]{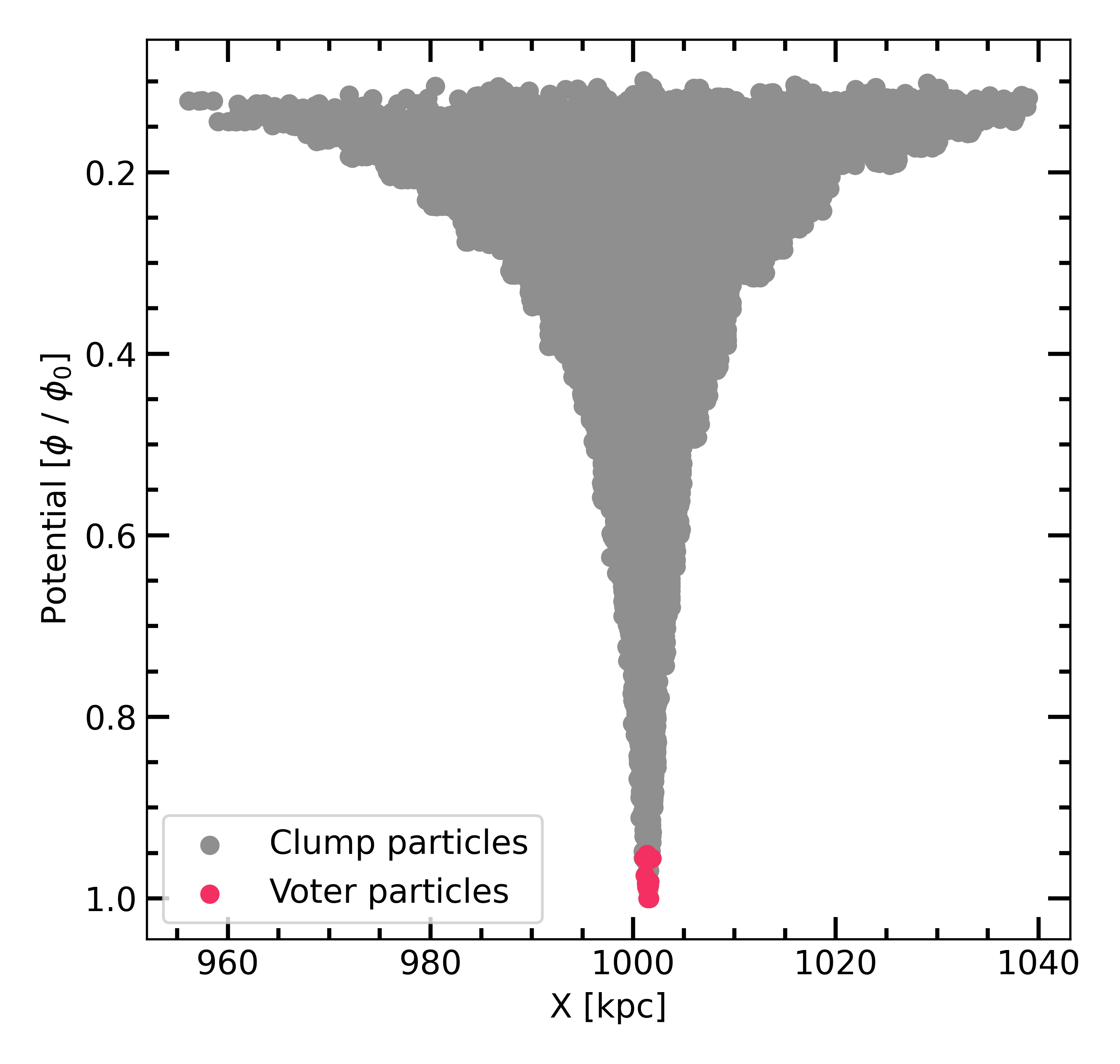} 
\centering
\caption{\label{fig:voters}
A projected position versus potential plot for a single isolated density peak. Here we show particles that belong to the peak in dark grey. The voter particles, identified by considering the kinetic energies and the potential displayed here, are shown in red.}
\end{figure}

There are various methods to connect structures across timesteps. Given our focus on tracking peaks instead of halos or sub-halos, we use particle membership information to link peaks across timesteps. The best particles for this task are the most bound particles within each structure because they tend to remain near the centre of the peak, where the gravitational potential is deepest (see Figure \ref{fig:voters}). These particles are also the most likely to endure interactions with other structures, making them ideal markers. We refer to these particles as ``voters" because their peak patch membership information is used to determine the peak's lineage. Each peak is assigned several voter particles whose memberships are assessed in the subsequent timestep. 

At each time step, each progenitor peak uses its most bound particles to ``vote" for a descendant by making a histogram of the voter particle's memberships at the descendant's timestep. The descendant that receives the majority of votes is then assigned to the progenitor. This method is powerful because the voters can vote for a descendant at any time step, not just adjacent ones, which allows for tracking peak evolution over extended periods and through non-consecutive steps. The voter system is also an improvement over the orphan particle system introduced in \cite{ACACIA}, as it eliminates the need to consider all particles in the peak patch to identify them. Lastly, a major advantage of this approach is that it is symmetric in time, which means the lineage of a peak can be traced in both directions of time. The details of tracking peaks across time steps are provided in Section \ref{subsec:Implementing_the_Merger_Tree_Algorithm}.

\subsection{Boosted Potential}
\label{subsec:boosted_pot}

\begin{figure*}
\includegraphics[width=\textwidth]{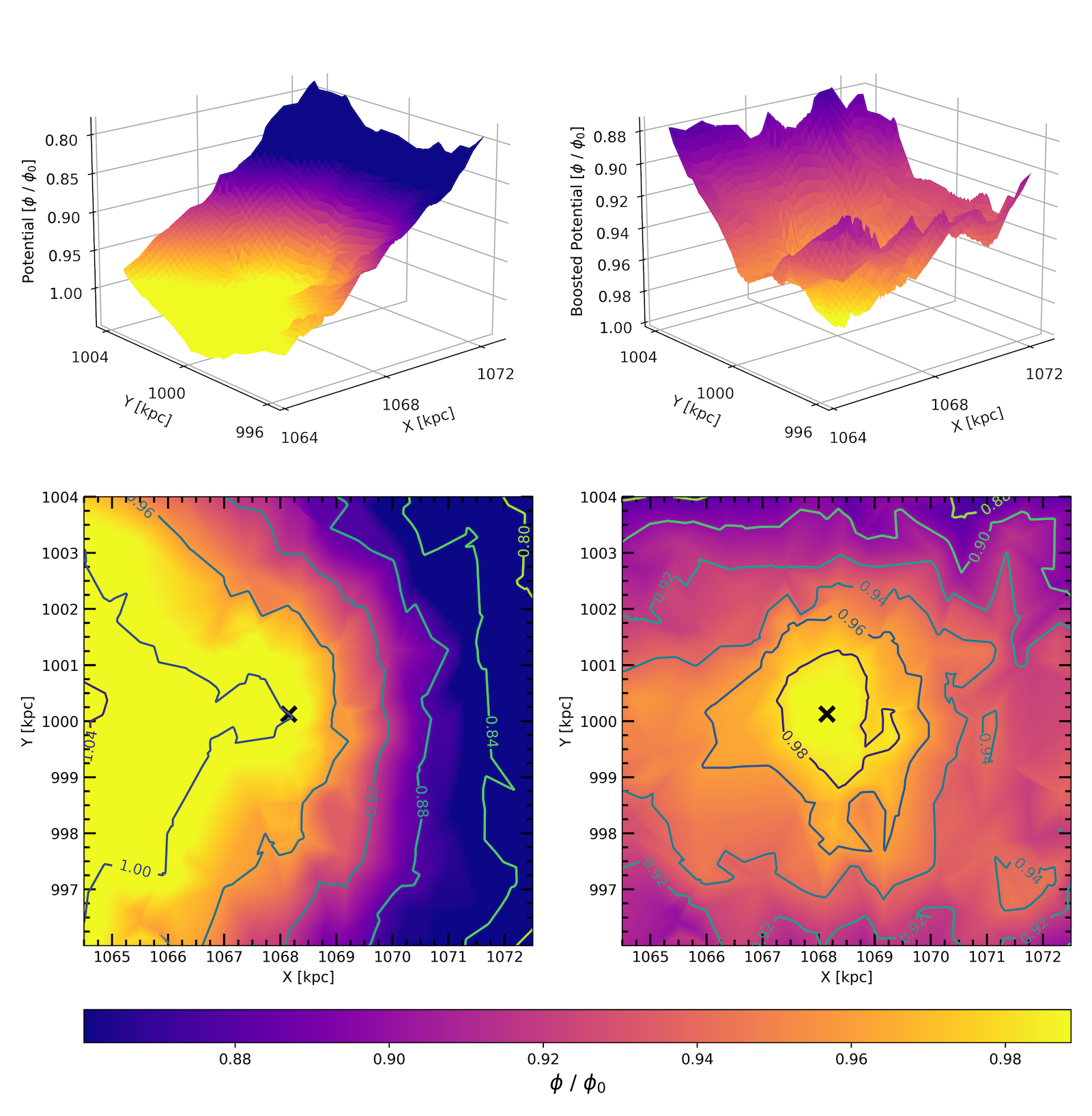} 
\centering
\caption{\label{fig:boosted}
A comparison of the boosted potential for a minor peak located near a more massive major peak. We show an x and y slice in each panel, with the minor peak located near the centre of the plots and the major peak, not shown, located to the left on the x-axis. The left panels show the potential landscape before subtracting out the gradient caused by the major peak, while the right panels show the boosted potential centred at the minor peak. The top panels show a 3D plot of the potential in each case, while the bottom panels show the heat map of the potential. For the heat maps, we include contours to show regions of equal potential, with $\Phi_0$ being the potential at the minor peak. Given the resolution and proximity of the minor peak to the major peak, the boosted potential offers a more meaningful potential landscape to approximate total energies relative to the minor peak.}
\end{figure*}

To define voters, we must first find the most bound particles within the identified peak patches. Gravitational potential defines gravitationally bound structures and is often used to verify whether particles are bound to these structures. Although the global potential ($\Phi({\bf x})$) is readily available, as it is directly computed during the simulation, it falls short compared to more advanced techniques because it lacks a clear correspondence to the underlying structures in the density field. In other words, a density maximum does not always correspond to a potential minimum. One commonly applied technique is the self-potential ($\Phi_{\rm self}({\bf x})$), which is calculated by considering only the local mass as the source of the potential and excluding external contributions ($\Phi_{\rm ext}({\bf x})$). However, this approach requires a good approximation of the local mass, or member particles of the structure, before the binding check. This can prove challenging in situations where structures, such as sub-halos, are embedded within large potential wells because they require careful treatment of large-scale tidal fields. For these reasons, we employ a third technique, the ``boosted potential'' introduced by \cite{boosted}, to define a locally meaningful potential landscape.

The boosted potential is defined by subtracting a uniform gradient from the global gravitational potential, which eliminates the effects of large-scale tidal fields. To construct the boosted potential, we assume that the large-scale gradient causes member particles of the structure in question to experience an approximately uniform acceleration, which \cite{boosted} refers to as the apparent acceleration ${\bf a_0}$. By incorporating ${\bf a_0}$, the accelerated, or  ``boosted", potential can be formulated as follows:
\begin{equation}
    \Phi_{\rm boost}({\bf x}) = \Phi({\bf x}) + {\bf a_0} \cdot {\bf x} 
\end{equation}
where ${\bf x_0}$ is the position of the peak and $\Phi({\bf x}) = \Phi_{\rm self}({\bf x}) + \Phi_{\rm ext}({\bf x})$. It can be also expressed as:
\begin{equation}
    \Phi_{\rm boost}({\bf x})= \Phi_{\rm self}({\bf x}) + \Phi_{\rm ext}({\bf x}) - ({\bf x - x_0} ) \frac{\partial \Phi_{\rm ext}}{\partial {\bf x}}({\bf x_0})
\label{eq:boosted}
\end{equation}

Where each potential contribution is more apparent. Once computed, $\Phi_{\rm boost}$ can serve as a more accurate recipe for computing the binding energy, analysing the disruption of sub-halos, and defining meaningful tidal boundaries. We refer the reader to \cite{boosted} for a detailed discussion on the boosted potential.

\subsection{Implementation and Voter Search}
\label{subsubsec:imp_boosted_pot}

For the binding energy ranking that we perform to find our $N_{\rm v}$ voters, we compute the kinetic and potential energies for all the candidate particles. For now, we define our candidate particles as all particles in the peak patch, but we will introduce selection improvements in Section \ref{subsubsec:optimizing_rad}. We compute the kinetic energies directly after subtracting out the bulk velocity of the peak that we have identified during the velocity deblending step. To compute the boosted potential, we must first estimate the apparent acceleration (${\bf a_0}$) due to the large-scale gradient. Given the assumption that the large-scale acceleration is approximately uniform, we compute ${\bf a_0}$ by taking the mean of the acceleration felt by all the candidate particles:

\begin{equation}
    {\bf a_0} = \frac{1}{N_c} \sum_{i=1}^{N_c} {\bf a_i}
\end{equation}

Where ${\bf a_i}$ is the acceleration of each candidate particle, and $N_c$ is the total number of such particles. As previously mentioned, the global potential is directly computed for each particle during the simulation. Using our estimated ${\bf a_0}$ and Equation \ref{eq:boosted}, we then compute the boosted potential for each of the candidate particles. We then add the kinetic energy to obtain the binding energy. We complete the voter search by finding the $N_{\rm v}$ most bound particles to the peak. 

\subsection{Optimizing the Search Radius}
\label{subsubsec:optimizing_rad}

\begin{figure}
\includegraphics[width=8cm]{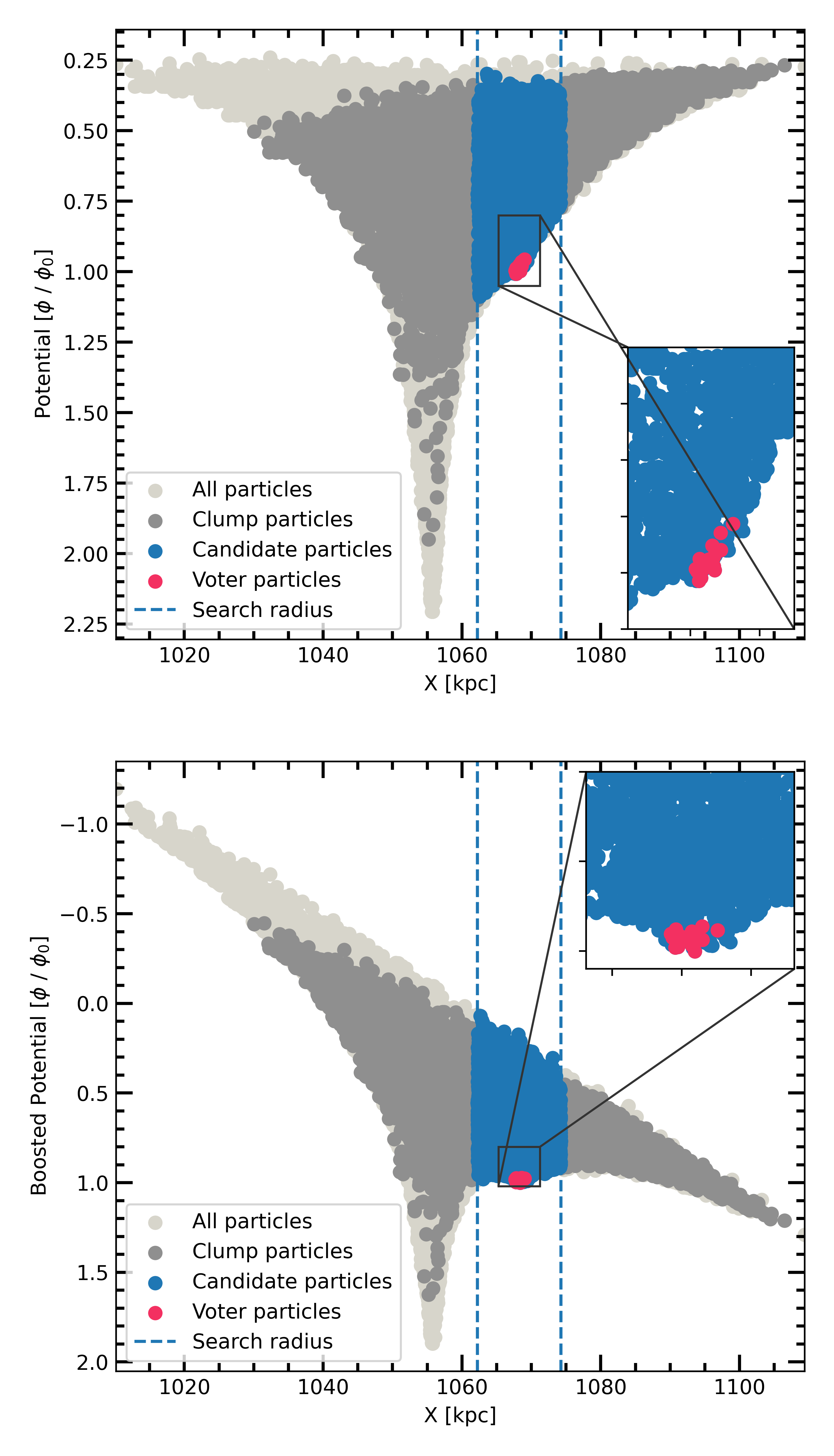} 
\centering
\caption{\label{fig:boosted_part}
A projected position versus potential plot for two overlapping density peaks from Figure \ref{fig:boosted}. The top panel shows the potential landscape without any corrections, while the bottom panel shows the landscape in an accelerated frame (i.e. the boosted potential). We use the same colours as Figure \ref{fig:voters} except we show particles that belong to the major peak in light grey and the particles that belong to the minor peak in dark grey. The zoomed-in portions of the plots also show that it is difficult to define binding energy without accounting for the large-scale gradient caused by the major peak.}
\end{figure}

\begin{figure*}
\includegraphics[width=\textwidth]{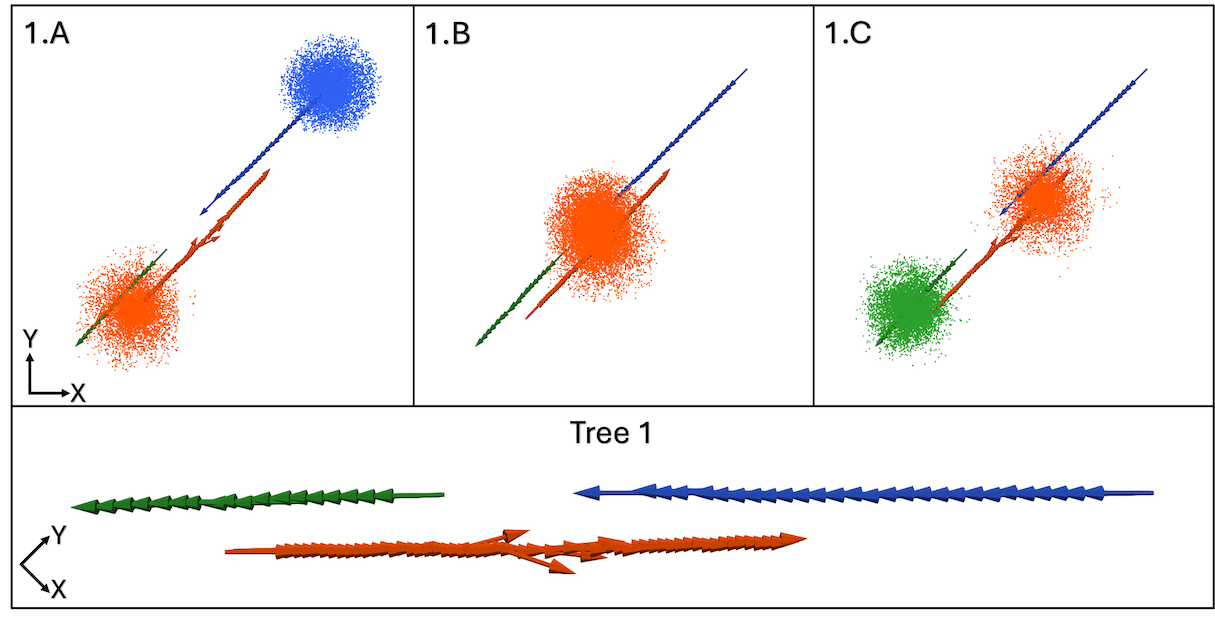} 
\centering
\caption{\label{fig:PHEW}
The top three panels display the major and minor peaks at three different timesteps: $5$, $35$, and $55$. Additionally, the path taken by each peak is shown as arrows pointing in the direction of the peak's velocity. These arrows' origins are positioned at the peak location at each timestep. We colour the peaks and particles according to the peak ID assigned by PHEW. The bottom panel shows this path as a way to represent the merger tree for the simulation. In panel 1.A we see that the peaks start off as two distinct peaks, blue and orange, but get confused as a single peak in panel 1.B. When the blue peak re-emerges, it is confused as a new peak and coloured green in the 1.C. Furthermore, we note that the velocity vector and position of the orange peak are affected by the confusion during overlap.}
\end{figure*}

As discussed in Section \ref{subsubsec:imp_boosted_pot}, careful selection of the particle sample from which we identify voters is essential, especially when a satellite peak is close to a dominant one. The boosted potential provides a good approximation of the satellite's local potential, but we must only consider particles within its tidal boundaries. If we do not constrain the search radius during our voter search, we will start to see particles with lower energies than the true voters due to their proximity to the dominant peak. \cite{boosted} addresses this issue by computing the critical contour that defines the tidal boundary of the peak patch. Though this is critical for defining and unbinding halos, our primary focus is identifying density peaks and their most bound particles. Consequently, we aim to avoid this additional computational overhead in favour of a simpler solution that does not require a topological assessment of the potential field. 

We address the issues outlined in this section by applying a spherical mask with radius $r_s$ around the density peak to identify voters. The initial search radius serves as a simulation-dependent maximum radius for voter searches and can be thought of as an ideal search radius that would enclose any isolated structure with a single embedded peak. This radius can be defined either as a global parameter or a dynamic value for each peak patch. For example, the peak-patch radius of the largest object expected in the simulation is a good approximation for a user-defined global parameter, while the distance to the most distant patch particle from the centre of each peak can estimate the value locally for each peak-patch. In this work, we start with a global value for the search radius, which we approximate to be half the size of the largest peak-patch we expect in the simulation.

For density peaks which have neighbouring peaks closer than the default search radius, we reduce the search radius based on the nearest significant saddle point in density. The saddle points are computed for substructures close to their dominant peaks during the PHEW peak-finding step \citep{PHEW, ACACIA}. In some cases, a saddle point is unavailable because the two clumps do not overlap despite their peaks being within each other's initial search radii. To address this issue, before each voter search we perform a neighbour proximity check. We first compute the distance to the nearest neighbour for each peak. We then choose the minimum value between the global search radius or one-fourth of the distance to the nearest neighbour. Though we do not implement this in this work, a minimum radius can also be defined to prevent the search radius from shrinking to sizes that do not contain any particles.

\subsection{Graveyard for Dead Peaks}

Keeping track of all the voter particles that were once members of density peaks but are now ``deceased" can take up a non-trivial amount of memory, especially if peaks frequently come in and out of existence. We, therefore, use a single array to store voter membership information when the peak they are associated with dissipates or merges (i.e. its lineage dies). Fittingly, we refer to the array as the ``graveyard" and use it to infer the continuity of peaks in non-subsequent timesteps. 

A graveyard array is essentially a one-dimensional integer array that matches the number of particles in a simulation. This array is initialized with all values set to a null value, such as $-1$ or $0$. It is worth mentioning that a sparse array representation can be utilized to conserve memory. Additionally, for simulations that do not assign unique peak IDs at each timestep, a second array can be employed to track the timestep of the merger. For the sake of clarity, we will focus on discussing a one-dimensional graveyard array as defined earlier.

As discussed in Section \ref{subsec:voter_sys}, each progenitor uses its voters to identify a descendant peak. If the progenitor does not identify a descendant, we assign the entries of the graveyard at the voters' indices to the peak ID of the progenitor. Should this peak resurface in subsequent timesteps without a clear progenitor from the previous timestep to claim it, it may initially appear to be a new peak. However, before labelling it as new, we use the voters of the re-emerging peak to vote for a progenitor using the membership information in the graveyard. The two are linked if a user-defined number of votes are given to the ``deceased" peak. This method, combined with the velocity deblending step, increases the reliability of the merger tree.

\subsection{Implementing the Merger Tree Algorithm}\label{subsec:Implementing_the_Merger_Tree_Algorithm}

This section outlines the detailed implementation of the merger tree algorithm. This algorithm can be applied during simulation runs for on-the-fly merger tree construction or as a post-processing step after the simulation run. In this section, we discuss on-the-fly merger tree implementation and provide the corresponding pseudocode in Appendix \ref{apdx:mta}. 

We assume that the peak finding step from Section \ref{sec:density_peak_finder} is complete for the current timestep (with or without velocity deblending), the voters of each peak are identified, and the peak catalogue from the last timestep is available. We refer to peaks in the current timestep as descendants and from the previous as progenitors. For cases where a descendant peak has a single progenitor, the progenitor is automatically labelled as the ``main progenitor". Lastly, a merger is registered when two or more progenitor peaks vote for the same descendant. This happens because the merging peaks are no longer identifiable within the surviving peak's patch. The question of which progenitor is the main progenitor is settled by having the descendant peak vote for a progenitor. Below we describe these steps in detail:

\begin{enumerate}
    \item For each peak, we compute the boosted potential and identify the $N_v$ most bound particles. We assign these particles as voters of their peak.
    \item We initiate a dictionary or hash table, \texttt{match\_dict}, for storing progenitor-descendant matches. The dictionary will have the descendant ID as its key and a linked list of possible progenitors as its value. We also initialize a list, \texttt{dead\_list}, to keep track of all the progenitors that failed to secure a descendant so we can add their voters to the graveyard at the end of the timestep.
    \item For each peak from the last step (progenitors), we look for the best possible descendant. We do so by using the progenitor's voters. Votes are cast by using the particle IDs of the voters to index the current timestep's peak patch membership array. We construct a histogram of the peak memberships.
    \begin{enumerate}
        \item If a descendant receives the most votes, the progenitors will be added to the \texttt{match\_dict}. Whether or not the progenitor is the main progenitor is checked in future steps. If noise in the voting is a concern, a parameter can be defined to constrain the minimum number of votes needed to constitute a match. 
        \item If the progenitor fails to secure any descendant candidates, it will be added to \texttt{dead\_list}. This could happen if the progenitor is dynamically destroyed or is unidentified by the peak finder. 
    \end{enumerate}
    \item We loop through all the unmatched descendants, which we can identify as descendants missing from \texttt{match\_dict}, to check for merger tree discontinuities. Similar to how votes are cast by the progenitors, each descendant casts its votes for a deceased peak using the graveyard array. We implement the graveyard, \texttt{grave\_yard},  as an array like the peak patch membership array, the difference being that the membership stored is the peak ID the voter was assigned at the time of the peak's death. If a deceased peak receives the user-parameterized number of votes, it and the voting descendant are added to \texttt{match\_dict}. Note that peak ID in the graveyard can be replaced if a particle is revived as a voter and its peak dies again. Furthermore, we allow for a maximum time limit, after which the graveyard will reset values for peaks that died beyond that limit, preventing them from reviving.

\begin{figure*}
\includegraphics[width=\textwidth]{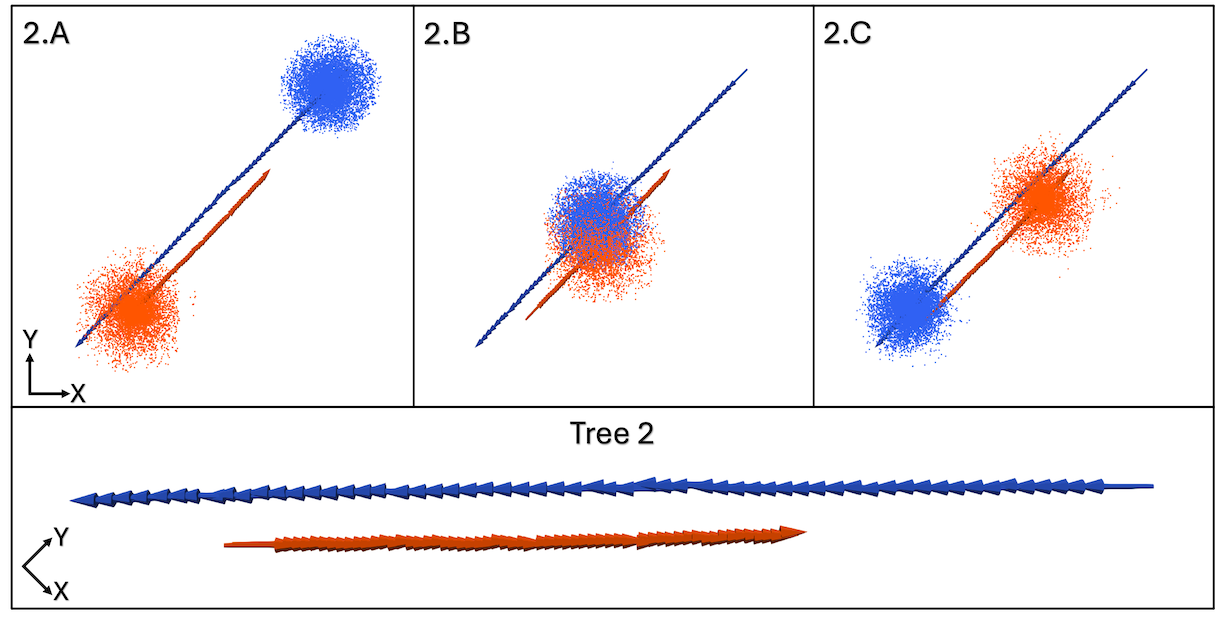} 
\centering
\caption{\label{fig:PHEW_voter}
Similar to Figure \ref{fig:PHEW}, the top three panels display the major and minor peaks at three different timesteps: $5$, $35$, and $55$. In this case, we colour the peaks and particles according to the peak ID assigned after PHEW segmentation and velocity deblending. The voting and graveyard systems are also activated for the merger tree construction, though they are unnecessary due to the velocity deblending in this case. This results in the deblending and identification of the two peaks during overlap, as shown in panel 2.B. Consequently, a consistent merger tree with accurate positions and bulk velocities is obtained.}
\end{figure*}

    \begin{enumerate}
        \item If a descendant remains unmatched after voting for a deceased peak, it suggests it is a newly formed peak. Before accepting this, we verify whether the peak was identified during the velocity deblending process. If this is the case, we flag the peak as an artefact because it violates our assertion that no new peak can form in another peak patch via velocity deblending. 
    \end{enumerate}
    \item At this stage, we possess a catalogue of matched progenitors for each descendant. We loop through these matches to determine the main progenitor for each descendant and flag potential mergers.
    \begin{enumerate}
        \item If there is only one progenitor candidate, mark that progenitor as the main progenitor. If there are multiple progenitors, there has been a merger in which one of the progenitors was absorbed into another's peak patch. In this case, we use the voters of the descendant and the previous timestep's peak patch membership array to identify the main progenitor. All other progenitors will be marked as merged with the descendant and added to \texttt{dead\_list}. The main progenitor and descendant are linked directly. 
    \end{enumerate}
    \item Lastly, we add the voters of all the deceased progenitors in \texttt{dead\_list} to the graveyard. In preparation for the next timestep, we save the current descendant as progenitors along with membership arrays. 
\end{enumerate}

\section{Demonstration}
\label{sec:Demonstration and Comparison}

In the previous sections, we introduced a phase-space peak finding algorithm and a corresponding merger tree algorithm to track the histories of these peaks. In this section, we demonstrate the algorithm outlined in Section~\ref{sec:density_peak_finder} and \ref{sec:merger_tree}. To evaluate the performance of our algorithm and its improvements, we conduct a series of tests involving a controlled collision of two dark matter halos. Each test is designed to highlight specific strengths and limitations of different methods for identifying and tracking structures. The results demonstrate how traditional spatial-only methods, phase-space enhancements, and our voting system address challenges such as overlapping structures and the accurate association of particles across time steps. In the following sections, we compare the performance of PHEW and ACACIA in Section \ref{demo:PHEW and ACACIA}, PHEW with voters in Section \ref{demo:PHEW and Voters Only}, and PHEW with velocity deblending in Section \ref{demo:PHEW, Velocity Deblending and Voters}, showcasing the incremental improvements in detecting and tracking these colliding halos.

\subsection{Initial Conditions and Simulation}

We initiate the simulation using DICE \citep{2016ascl.soft07002P} by configuring two virialized dark matter halos, which we will refer to as the ``major" and ``minor" halos. Both halos are initialized with $10,000$ particles each and separated by about $241$ kpc. The major halo is given a virial mass of $10^{10}\ {\rm M_\odot}$ and placed at the origin of the simulation. Furthermore, it is initialized with a zero bulk velocity. The second halo, the minor halo, is initialized with a virial mass of $5 \times 10^9\ {\rm M_\odot}$  at a position of $(160, 180, 0)$ kpc, moving towards the first galaxy with a bulk velocity of $(-353, -353, 0)$ km/s. This configuration is specifically chosen to simulate a collision trajectory between the two halos with velocities well above each halo's escape velocity. The concentration of the minor halo is deliberately set to a low value, $c=2$, to create a challenging scenario for its detection and mimic tidally disrupted halos. We start the simulation and integrate for $447$ million years; writing outputs every $7.5$ million year ($60$ time-steps).

\begin{table}
\centering
\begin{tabular}{l*{3}{l}}
\hline
\hline
& Property & Major & Minor \\ 
\hline
& Number of particles & $10^{4}$ & $10^{4}$ \\ 
& Mass [$10^{10} {\rm M_\odot}$] & $1$ & $0.5$ \\ 
& Concentration & $20$ & $2$ \\ 
& Position [kpc] & $(0, 0, 0)$ & $(160, 180, 0)$ \\ 
& Velocity [km/s] & $(0, 0, 0)$ & $(-353, -353, 0)$ \\ 
\hline
\end{tabular}
\caption{NFW parameters of the major and minor halos used in this demonstration.}\label{table:nfw}
\end{table}

\subsection{PHEW and ACACIA}
\label{demo:PHEW and ACACIA}

We use the PHEW clump finder and the ACACIA merger tree code to analyse the path of the colliding dark matter halos. The PHEW clump finder is adept at detecting density peaks and their associated peak-patches both before and after the halos overlap spatially. However, we set the relevance threshold to the very high value of $100$, such that PHEW directly merges the two halos when their peak-patches significantly overlap in space. The relevance threshold determines the minimum ratio in density between a peak and its surrounding saddle points for the peak to be considered significant. By choosing such a high value, we effectively merge halos whose density profiles significantly overlap, simplifying the analysis to focus on the broader challenges of overlapping structures. In this case, since PHEW does not utilize phase-space information, it struggles to differentiate between the overlapping halos. This limitation of the clump finder affects the performance of the ACACIA merger tree algorithm. ACACIA, which relies on the input from PHEW, erroneously identifies the minor halo as two separate entities before and after its overlap with the major halo. Figure \ref{fig:PHEW} shows three snapshots and the path taken by each peak as determined by PHEW and ACACIA. 

\subsection{PHEW and Voters Only}
\label{demo:PHEW and Voters Only}

We take the same PHEW segmentation and use the voting system to construct the merger tree. We find $20$ voters for each of the clumps PHEW identified, using the boosted potential with a search radius of $20$ kpc. Choosing more voters offers a wider view of particle membership, but too many voters could bias the votes as less bound particles wander out of the peak. In contrast, though fewer voters will almost always vote in favour of the correct peak, choosing too few may result in the under-sampling of the memberships of the most bound particles. Furthermore, we required $15$ votes for a candidate to win during the descendant or progenitor selection step and $10$ votes for the graveyard vote for a deceased progenitor to be reactivated. As expected, the clumps are misidentified as a single clump multiple times during the steps they overlap. But each time the clumps reemerged as separate, the voter system and the graveyard array correctly identified the progenitors. It's also good to note that the positions of the voters could be used to trace the path of the confused peaks in a post-processing step. 

\subsection{PHEW, Velocity Deblending and Voters}
\label{demo:PHEW, Velocity Deblending and Voters}

Lastly, we take the PHEW spatial segmentation and apply the velocity deblending step to each of the identified clumps. We then find $20$ voters for each of the clumps using the boosted potential. We define the search radius again as $20$ kpc. Using the voting system and the graveyard, we construct a merger tree. Given the large velocity difference between the two peaks, the deblending step separates the two peaks, even when they are overlapping. Despite being active, the graveyard was not used because none of the peaks failed to secure a descendant. Figure \ref{fig:PHEW_voter} shows three snapshots and the path taken by each peak as determined by this method, which is a vast improvement over Figure \ref{fig:PHEW}, which shows the PHEW and ACACIA method. It is worth noting that several artefacts appeared during the velocity deblending step but were correctly identified as such during the merger tree step.  

This paper presents a new watershed algorithm that uses phase-space information to identify structures, especially in complex environments where traditional methods may struggle due to spatially overlapping structures. We include a merger tree algorithm to track density peaks across time steps, which uses the boosted potential to identify the most bound particles for each density peak. The key contributions and concepts presented in this work can be summarized as follows:

\begin{enumerate}
\item A density peak, as defined in the context of simulations, can be parameterized by seven coordinates: six phase-space coordinates and one time coordinate, which we denote as the ``peak coordinates”. 

\item In addition to the density peak coordinates, it is necessary for each peak to be associated with a few most bound particles for tracking. We call these most bound particles the peak's ``voters" because their membership is used to vote for a progenitor or descendant. 

\item We introduced velocity deblending to deblend overlapping density peaks, improving our ability to track cosmic structure where traditional spatial-only methods fail.

\item Our voter system leverages the most bound particles within structures to trace peak lineages across timesteps, even when those timesteps are not sequential.

\item We applied the concept of boosted potentials to improve the identification of gravitationally bound particles, which account for large-scale tidal effects, improving voter selection and the accuracy of our merger trees.
\end{enumerate}

The work presented in this paper is a significant step towards the development of a reliable structure finder and merger tree code. Our main goal is to derive the simplest algorithm and definition of structure possible while maintaining or improving reliability. In future work, we aim to use our definition of structure and the methods we discussed to construct catalogues that we will compare to observational data. This comparison has the potential to significantly advance our understanding of cosmic structures. We also plan to test and compare the effectiveness of our algorithm using large-scale and zoom-in cosmological simulations, which could further validate our approach and its applicability.

\section*{Acknowledgements}

The authors are pleased to acknowledge that the work reported on in this paper was substantially performed using the Princeton Research Computing resources at Princeton University which is consortium of groups led by the Princeton Institute for Computational Science and Engineering (PICSciE) and Office of Information Technology's Research Computing. This material is based upon work supported by the National Aeronautics and Space Administration under Grant No. FP00005620 issued through the Jet Propulsion Laboratory (JPL). This work made use of the following software packages: \texttt{astropy} \citep{astropy:2013, astropy:2018, astropy:2022}, \texttt{matplotlib} \citep{Hunter:2007}, \texttt{numpy} \citep{numpy}, \texttt{python} \citep{python}, \texttt{scipy} \citep{2020SciPy-NMeth, scipy_11702230}, \texttt{DICE} \citep{2016ascl.soft07002P} and \texttt{RAMSES} \citep{2002A&A...385..337T}.

Software citation information aggregated using \texttt{\href{https://www.tomwagg.com/software-citation-station/}{The Software Citation Station}} \citep{software-citation-station-paper, software-citation-station-zenodo}.

\section*{Data Availability}
The simulation data used for this work will be provided upon reasonable request to the corresponding author. 




\bibliographystyle{mnras}
\bibliography{RHF} 

\begin{thebibliography}{}
\makeatletter
\relax
\def\mn@urlcharsother{\let\do\@makeother \do\$\do\&\do\#\do\^\do\_\do\%\do\~}
\def\mn@doi{\begingroup\mn@urlcharsother \@ifnextchar [ {\mn@doi@} {\mn@doi@[]}}
\def\mn@doi@[#1]#2{\def\@tempa{#1}\ifx\@tempa\@empty \href {http://dx.doi.org/#2} {doi:#2}\else \href {http://dx.doi.org/#2} {#1}\fi \endgroup}
\def\mn@eprint#1#2{\mn@eprint@#1:#2::\@nil}
\def\mn@eprint@arXiv#1{\href {http://arxiv.org/abs/#1} {{\tt arXiv:#1}}}
\def\mn@eprint@dblp#1{\href {http://dblp.uni-trier.de/rec/bibtex/#1.xml} {dblp:#1}}
\def\mn@eprint@#1:#2:#3:#4\@nil{\def\@tempa {#1}\def\@tempb {#2}\def\@tempc {#3}\ifx \@tempc \@empty \let \@tempc \@tempb \let \@tempb \@tempa \fi \ifx \@tempb \@empty \def\@tempb {arXiv}\fi \@ifundefined {mn@eprint@\@tempb}{\@tempb:\@tempc}{\expandafter \expandafter \csname mn@eprint@\@tempb\endcsname \expandafter{\@tempc}}}

\bibitem[\protect\citeauthoryear{{Abbott} et~al.,}{{Abbott} et~al.}{2018}]{2018PhRvD..98d3526A}
{Abbott} T.~M.~C.,  et~al., 2018, \mn@doi [\prd] {10.1103/PhysRevD.98.043526}, \href {https://ui.adsabs.harvard.edu/abs/2018PhRvD..98d3526A} {98, 043526}

\bibitem[\protect\citeauthoryear{{Adhikari}, {Dalal}  \& {Chamberlain}}{{Adhikari} et~al.}{2014}]{2014JCAP...11..019A}
{Adhikari} S.,  {Dalal} N.,   {Chamberlain} R.~T.,  2014, \mn@doi [\jcap] {10.1088/1475-7516/2014/11/019}, \href {https://ui.adsabs.harvard.edu/abs/2014JCAP...11..019A} {2014, 019}

\bibitem[\protect\citeauthoryear{{Ascasibar} \& {Gottl{\"o}ber}}{{Ascasibar} \& {Gottl{\"o}ber}}{2008}]{2008MNRAS.386.2022A}
{Ascasibar} Y.,  {Gottl{\"o}ber} S.,  2008, \mn@doi [\mnras] {10.1111/j.1365-2966.2008.13160.x}, \href {https://ui.adsabs.harvard.edu/abs/2008MNRAS.386.2022A} {386, 2022}

\bibitem[\protect\citeauthoryear{{Astropy Collaboration} et~al.,}{{Astropy Collaboration} et~al.}{2013}]{astropy:2013}
{Astropy Collaboration} et~al., 2013, \mn@doi [\aap] {10.1051/0004-6361/201322068}, \href {http://adsabs.harvard.edu/abs/2013A%26A...558A..33A} {558, A33}

\bibitem[\protect\citeauthoryear{{Astropy Collaboration} et~al.,}{{Astropy Collaboration} et~al.}{2018}]{astropy:2018}
{Astropy Collaboration} et~al., 2018, \mn@doi [\aj] {10.3847/1538-3881/aabc4f}, \href {https://ui.adsabs.harvard.edu/abs/2018AJ....156..123A} {156, 123}

\bibitem[\protect\citeauthoryear{{Astropy Collaboration} et~al.,}{{Astropy Collaboration} et~al.}{2022}]{astropy:2022}
{Astropy Collaboration} et~al., 2022, \mn@doi [\apj] {10.3847/1538-4357/ac7c74}, \href {https://ui.adsabs.harvard.edu/abs/2022ApJ...935..167A} {935, 167}

\bibitem[\protect\citeauthoryear{{Behroozi}, {Wechsler}  \& {Wu}}{{Behroozi} et~al.}{2013}]{2013ApJ...762..109B}
{Behroozi} P.~S.,  {Wechsler} R.~H.,   {Wu} H.-Y.,  2013, \mn@doi [\apj] {10.1088/0004-637X/762/2/109}, \href {https://ui.adsabs.harvard.edu/abs/2013ApJ...762..109B} {762, 109}

\bibitem[\protect\citeauthoryear{{Bleuler} \& {Teyssier}}{{Bleuler} \& {Teyssier}}{2014}]{PHEW_Old}
{Bleuler} A.,  {Teyssier} R.,  2014, \mn@doi [\mnras] {10.1093/mnras/stu2005}, \href {https://ui.adsabs.harvard.edu/abs/2014MNRAS.445.4015B} {445, 4015}

\bibitem[\protect\citeauthoryear{{Bleuler}, {Teyssier}, {Carassou}  \& {Martizzi}}{{Bleuler} et~al.}{2015}]{PHEW}
{Bleuler} A.,  {Teyssier} R.,  {Carassou} S.,   {Martizzi} D.,  2015, \mn@doi [Computational Astrophysics and Cosmology] {10.1186/s40668-015-0009-7}, \href {https://ui.adsabs.harvard.edu/abs/2015ComAC...2....5B} {2, 5}

\bibitem[\protect\citeauthoryear{{Booth} \& {Schaye}}{{Booth} \& {Schaye}}{2009}]{2009MNRAS.398...53B}
{Booth} C.~M.,  {Schaye} J.,  2009, \mn@doi [\mnras] {10.1111/j.1365-2966.2009.15043.x}, \href {https://ui.adsabs.harvard.edu/abs/2009MNRAS.398...53B} {398, 53}

\bibitem[\protect\citeauthoryear{{Davis}, {Efstathiou}, {Frenk}  \& {White}}{{Davis} et~al.}{1985}]{1985ApJ...292..371D}
{Davis} M.,  {Efstathiou} G.,  {Frenk} C.~S.,   {White} S.~D.~M.,  1985, \mn@doi [\apj] {10.1086/163168}, \href {https://ui.adsabs.harvard.edu/abs/1985ApJ...292..371D} {292, 371}

\bibitem[\protect\citeauthoryear{{Diemer} \& {Kravtsov}}{{Diemer} \& {Kravtsov}}{2014}]{2014ApJ...789....1D}
{Diemer} B.,  {Kravtsov} A.~V.,  2014, \mn@doi [\apj] {10.1088/0004-637X/789/1/1}, \href {https://ui.adsabs.harvard.edu/abs/2014ApJ...789....1D} {789, 1}

\bibitem[\protect\citeauthoryear{{Eisenstein} \& {Hut}}{{Eisenstein} \& {Hut}}{1998}]{1998ApJ...498..137E}
{Eisenstein} D.~J.,  {Hut} P.,  1998, \mn@doi [\apj] {10.1086/305535}, \href {https://ui.adsabs.harvard.edu/abs/1998ApJ...498..137E} {498, 137}

\bibitem[\protect\citeauthoryear{{Fumagalli}, {Costanzi}, {Saro}, {Castro}  \& {Borgani}}{{Fumagalli} et~al.}{2024}]{2024A&A...682A.148F}
{Fumagalli} A.,  {Costanzi} M.,  {Saro} A.,  {Castro} T.,   {Borgani} S.,  2024, \mn@doi [\aap] {10.1051/0004-6361/202348296}, \href {https://ui.adsabs.harvard.edu/abs/2024A&A...682A.148F} {682, A148}

\bibitem[\protect\citeauthoryear{{Gao}, {Frenk}, {Boylan-Kolchin}, {Jenkins}, {Springel}  \& {White}}{{Gao} et~al.}{2011}]{2011MNRAS.410.2309G}
{Gao} L.,  {Frenk} C.~S.,  {Boylan-Kolchin} M.,  {Jenkins} A.,  {Springel} V.,   {White} S.~D.~M.,  2011, \mn@doi [\mnras] {10.1111/j.1365-2966.2010.17601.x}, \href {https://ui.adsabs.harvard.edu/abs/2011MNRAS.410.2309G} {410, 2309}

\bibitem[\protect\citeauthoryear{{Ghigna}, {Moore}, {Governato}, {Lake}, {Quinn}  \& {Stadel}}{{Ghigna} et~al.}{2000}]{2000ApJ...544..616G}
{Ghigna} S.,  {Moore} B.,  {Governato} F.,  {Lake} G.,  {Quinn} T.,   {Stadel} J.,  2000, \mn@doi [\apj] {10.1086/317221}, \href {https://ui.adsabs.harvard.edu/abs/2000ApJ...544..616G} {544, 616}

\bibitem[\protect\citeauthoryear{{Gill}, {Knebe}  \& {Gibson}}{{Gill} et~al.}{2004}]{2004MNRAS.351..399G}
{Gill} S. P.~D.,  {Knebe} A.,   {Gibson} B.~K.,  2004, \mn@doi [\mnras] {10.1111/j.1365-2966.2004.07786.x}, \href {https://ui.adsabs.harvard.edu/abs/2004MNRAS.351..399G} {351, 399}

\bibitem[\protect\citeauthoryear{Gommers et~al.,}{Gommers et~al.}{2024}]{scipy_11702230}
Gommers R.,  et~al., 2024, scipy/scipy: SciPy 1.14.0rc2, \mn@doi{10.5281/zenodo.11702230}, \url {https://doi.org/10.5281/zenodo.11702230}

\bibitem[\protect\citeauthoryear{{Hadzhiyska}, {Eisenstein}, {Bose}, {Garrison}  \& {Maksimova}}{{Hadzhiyska} et~al.}{2022}]{2022MNRAS.509..501H}
{Hadzhiyska} B.,  {Eisenstein} D.,  {Bose} S.,  {Garrison} L.~H.,   {Maksimova} N.,  2022, \mn@doi [\mnras] {10.1093/mnras/stab2980}, \href {https://ui.adsabs.harvard.edu/abs/2022MNRAS.509..501H} {509, 501}

\bibitem[\protect\citeauthoryear{Harris et~al.,}{Harris et~al.}{2020}]{numpy}
Harris C.~R.,  et~al., 2020, \mn@doi [Nature] {10.1038/s41586-020-2649-2}, 585, 357

\bibitem[\protect\citeauthoryear{Hunter}{Hunter}{2007}]{Hunter:2007}
Hunter J.~D.,  2007, \mn@doi [Computing in Science \& Engineering] {10.1109/MCSE.2007.55}, 9, 90

\bibitem[\protect\citeauthoryear{{Ivkovic} \& {Teyssier}}{{Ivkovic} \& {Teyssier}}{2022}]{ACACIA}
{Ivkovic} M.,  {Teyssier} R.,  2022, \mn@doi [\mnras] {10.1093/mnras/stab3329}, \href {https://ui.adsabs.harvard.edu/abs/2022MNRAS.510..959I} {510, 959}

\bibitem[\protect\citeauthoryear{{Klypin}, {Gottl{\"o}ber}, {Kravtsov}  \& {Khokhlov}}{{Klypin} et~al.}{1999}]{1999ApJ...516..530K}
{Klypin} A.,  {Gottl{\"o}ber} S.,  {Kravtsov} A.~V.,   {Khokhlov} A.~M.,  1999, \mn@doi [\apj] {10.1086/307122}, \href {https://ui.adsabs.harvard.edu/abs/1999ApJ...516..530K} {516, 530}

\bibitem[\protect\citeauthoryear{{Knebe} et~al.,}{{Knebe} et~al.}{2011}]{2011MNRAS.415.2293K}
{Knebe} A.,  et~al., 2011, \mn@doi [\mnras] {10.1111/j.1365-2966.2011.18858.x}, \href {https://ui.adsabs.harvard.edu/abs/2011MNRAS.415.2293K} {415, 2293}

\bibitem[\protect\citeauthoryear{{Knollmann} \& {Knebe}}{{Knollmann} \& {Knebe}}{2009}]{2009ApJS..182..608K}
{Knollmann} S.~R.,  {Knebe} A.,  2009, \mn@doi [\apjs] {10.1088/0067-0049/182/2/608}, \href {https://ui.adsabs.harvard.edu/abs/2009ApJS..182..608K} {182, 608}

\bibitem[\protect\citeauthoryear{{Lacey} \& {Cole}}{{Lacey} \& {Cole}}{1994}]{1994MNRAS.271..676L}
{Lacey} C.,  {Cole} S.,  1994, \mn@doi [\mnras] {10.1093/mnras/271.3.676}, \href {https://ui.adsabs.harvard.edu/abs/1994MNRAS.271..676L} {271, 676}

\bibitem[\protect\citeauthoryear{{Ma}, {Takeuchi}, {Cooray}  \& {Zhu}}{{Ma} et~al.}{2024}]{2024arXiv240509855M}
{Ma} H.-X.,  {Takeuchi} T.~T.,  {Cooray} S.,   {Zhu} Y.,  2024, \mn@doi [arXiv e-prints] {10.48550/arXiv.2405.09855}, \href {https://ui.adsabs.harvard.edu/abs/2024arXiv240509855M} {p. arXiv:2405.09855}

\bibitem[\protect\citeauthoryear{{Moore}, {Katz}  \& {Lake}}{{Moore} et~al.}{1996}]{1996ApJ...457..455M}
{Moore} B.,  {Katz} N.,   {Lake} G.,  1996, \mn@doi [\apj] {10.1086/176745}, \href {https://ui.adsabs.harvard.edu/abs/1996ApJ...457..455M} {457, 455}

\bibitem[\protect\citeauthoryear{{More}, {Diemer}  \& {Kravtsov}}{{More} et~al.}{2015}]{2015ApJ...810...36M}
{More} S.,  {Diemer} B.,   {Kravtsov} A.~V.,  2015, \mn@doi [\apj] {10.1088/0004-637X/810/1/36}, \href {https://ui.adsabs.harvard.edu/abs/2015ApJ...810...36M} {810, 36}

\bibitem[\protect\citeauthoryear{{Navarro}, {Frenk}  \& {White}}{{Navarro} et~al.}{1996}]{1996ApJ...462..563N}
{Navarro} J.~F.,  {Frenk} C.~S.,   {White} S. D.~M.,  1996, \mn@doi [\apj] {10.1086/177173}, \href {https://ui.adsabs.harvard.edu/abs/1996ApJ...462..563N} {462, 563}

\bibitem[\protect\citeauthoryear{{Perret}}{{Perret}}{2016}]{2016ascl.soft07002P}
{Perret} V.,  2016, {DICE: Disk Initial Conditions Environment}, Astrophysics Source Code Library, record ascl:1607.002

\bibitem[\protect\citeauthoryear{{Press} \& {Schechter}}{{Press} \& {Schechter}}{1974}]{1974ApJ...187..425P}
{Press} W.~H.,  {Schechter} P.,  1974, \mn@doi [\apj] {10.1086/152650}, \href {https://ui.adsabs.harvard.edu/abs/1974ApJ...187..425P} {187, 425}

\bibitem[\protect\citeauthoryear{{Sijacki}, {Springel}, {Di Matteo}  \& {Hernquist}}{{Sijacki} et~al.}{2007}]{2007MNRAS.380..877S}
{Sijacki} D.,  {Springel} V.,  {Di Matteo} T.,   {Hernquist} L.,  2007, \mn@doi [\mnras] {10.1111/j.1365-2966.2007.12153.x}, \href {https://ui.adsabs.harvard.edu/abs/2007MNRAS.380..877S} {380, 877}

\bibitem[\protect\citeauthoryear{{Springel}, {White}, {Tormen}  \& {Kauffmann}}{{Springel} et~al.}{2001}]{2001MNRAS.328..726S}
{Springel} V.,  {White} S. D.~M.,  {Tormen} G.,   {Kauffmann} G.,  2001, \mn@doi [\mnras] {10.1046/j.1365-8711.2001.04912.x}, \href {https://ui.adsabs.harvard.edu/abs/2001MNRAS.328..726S} {328, 726}

\bibitem[\protect\citeauthoryear{{Springel}, {Di Matteo}  \& {Hernquist}}{{Springel} et~al.}{2005}]{2005MNRAS.361..776S}
{Springel} V.,  {Di Matteo} T.,   {Hernquist} L.,  2005, \mn@doi [\mnras] {10.1111/j.1365-2966.2005.09238.x}, \href {https://ui.adsabs.harvard.edu/abs/2005MNRAS.361..776S} {361, 776}

\bibitem[\protect\citeauthoryear{{St{\"u}cker}, {Angulo}  \& {Busch}}{{St{\"u}cker} et~al.}{2021}]{boosted}
{St{\"u}cker} J.,  {Angulo} R.~E.,   {Busch} P.,  2021, \mn@doi [\mnras] {10.1093/mnras/stab2913}, \href {https://ui.adsabs.harvard.edu/abs/2021MNRAS.508.5196S} {508, 5196}

\bibitem[\protect\citeauthoryear{{Suhhonenko} \& {Gramann}}{{Suhhonenko} \& {Gramann}}{2003}]{2003MNRAS.339..271S}
{Suhhonenko} I.,  {Gramann} M.,  2003, \mn@doi [\mnras] {10.1046/j.1365-8711.2003.06175.x}, \href {https://ui.adsabs.harvard.edu/abs/2003MNRAS.339..271S} {339, 271}

\bibitem[\protect\citeauthoryear{{Taylor} \& {Kobayashi}}{{Taylor} \& {Kobayashi}}{2014}]{2014MNRAS.442.2751T}
{Taylor} P.,  {Kobayashi} C.,  2014, \mn@doi [\mnras] {10.1093/mnras/stu983}, \href {https://ui.adsabs.harvard.edu/abs/2014MNRAS.442.2751T} {442, 2751}

\bibitem[\protect\citeauthoryear{{Teyssier}}{{Teyssier}}{2002}]{2002A&A...385..337T}
{Teyssier} R.,  2002, \mn@doi [\aap] {10.1051/0004-6361:20011817}, \href {https://ui.adsabs.harvard.edu/abs/2002A&A...385..337T} {385, 337}

\bibitem[\protect\citeauthoryear{{Tormen}, {Diaferio}  \& {Syer}}{{Tormen} et~al.}{1998}]{1998MNRAS.299..728T}
{Tormen} G.,  {Diaferio} A.,   {Syer} D.,  1998, \mn@doi [\mnras] {10.1046/j.1365-8711.1998.01775.x}, \href {https://ui.adsabs.harvard.edu/abs/1998MNRAS.299..728T} {299, 728}

\bibitem[\protect\citeauthoryear{{Tremmel}, {Karcher}, {Governato}, {Volonteri}, {Quinn}, {Pontzen}, {Anderson}  \& {Bellovary}}{{Tremmel} et~al.}{2017}]{2017MNRAS.470.1121T}
{Tremmel} M.,  {Karcher} M.,  {Governato} F.,  {Volonteri} M.,  {Quinn} T.~R.,  {Pontzen} A.,  {Anderson} L.,   {Bellovary} J.,  2017, \mn@doi [\mnras] {10.1093/mnras/stx1160}, \href {https://ui.adsabs.harvard.edu/abs/2017MNRAS.470.1121T} {470, 1121}

\bibitem[\protect\citeauthoryear{Van~Rossum \& Drake}{Van~Rossum \& Drake}{2009}]{python}
Van~Rossum G.,  Drake F.~L.,  2009, Python 3 Reference Manual.
CreateSpace, Scotts Valley, CA

\bibitem[\protect\citeauthoryear{Virtanen et~al.,}{Virtanen et~al.}{2020}]{2020SciPy-NMeth}
Virtanen P.,  et~al., 2020, \mn@doi [Nature Methods] {10.1038/s41592-019-0686-2}, \href {https://rdcu.be/b08Wh} {17, 261}

\bibitem[\protect\citeauthoryear{Wagg \& Broekgaarden}{Wagg \& Broekgaarden}{2024a}]{software-citation-station-zenodo}
Wagg T.,  Broekgaarden F.,  2024a, The Software Citation Station, \mn@doi{10.5281/zenodo.11292917}, \url {https://doi.org/10.5281/zenodo.11292917}

\bibitem[\protect\citeauthoryear{{Wagg} \& {Broekgaarden}}{{Wagg} \& {Broekgaarden}}{2024b}]{software-citation-station-paper}
{Wagg} T.,  {Broekgaarden} F.~S.,  2024b, arXiv e-prints, \href {https://ui.adsabs.harvard.edu/abs/2024arXiv240604405W} {p. arXiv:2406.04405}

\bibitem[\protect\citeauthoryear{{Warren}, {Quinn}, {Salmon}  \& {Zurek}}{{Warren} et~al.}{1992}]{1992ApJ...399..405W}
{Warren} M.~S.,  {Quinn} P.~J.,  {Salmon} J.~K.,   {Zurek} W.~H.,  1992, \mn@doi [\apj] {10.1086/171937}, \href {https://ui.adsabs.harvard.edu/abs/1992ApJ...399..405W} {399, 405}

\bibitem[\protect\citeauthoryear{{White}, {Davis}, {Efstathiou}  \& {Frenk}}{{White} et~al.}{1987}]{1987Natur.330..451W}
{White} S. D.~M.,  {Davis} M.,  {Efstathiou} G.,   {Frenk} C.~S.,  1987, \mn@doi [\nat] {10.1038/330451a0}, \href {https://ui.adsabs.harvard.edu/abs/1987Natur.330..451W} {330, 451}

\bibitem[\protect\citeauthoryear{{van Kampen}, {Jimenez}  \& {Peacock}}{{van Kampen} et~al.}{1999}]{1999MNRAS.310...43V}
{van Kampen} E.,  {Jimenez} R.,   {Peacock} J.~A.,  1999, \mn@doi [\mnras] {10.1046/j.1365-8711.1999.02955.x}, \href {https://ui.adsabs.harvard.edu/abs/1999MNRAS.310...43V} {310, 43}

\makeatother
\end{thebibliography}




\clearpage

\appendix

\section{Merger Tree Algorithm}
\label{apdx:mta}

Here, we provide a pseudocode of the merger tree algorithm. This algorithm can be run either at the end of each time step or as a post-processing step. This merger tree code assumes that density peaks have been found and the velocity deblending step has been applied.

\begin{algorithm}[]
    \SetAlgoLined
    \For{each timestep}{   
        \For{descendant from this timestep}{
            descendant.boosted\_potential(r\_search)\;
            descendant.find\_voters()\;
        }
        \BlankLine
        \BlankLine
        match\_dict = \{\}\;
        dead\_list = [ ]\;
        \For{progenitor from last timestep}{
            \BlankLine
            \If{progenitor.is\_artefact}{skip\;}
            \BlankLine
            progenitor.vote(current\_particle\_membership)\;
            \eIf{descendant is found}{
                match\_dict[descendant].add(progenitor)\;
                }{
                dead\_list.add(progenitor)\;
            }
        }
        \BlankLine
        \BlankLine
        \For{unmatched descendant}{
            \BlankLine
            descendant.vote(grave\_yard)\;
            \BlankLine
            \If{deceased progenitor is found}{
                match\_dict[descendant].add(dead progenitor)\;
            }
            \BlankLine
            \If{descendant is still unmatched}{
                \If{descendant found in velocity deblending step}{
                    descendant.is\_artefact = True \;
                }
            }
        }
        \BlankLine
        \BlankLine
        \For{match in  match\_dict}{
            \BlankLine
            \eIf{single progenitor}{
                main\_progenitor = single\_progenitor
            }{
                main\_progenitor =  descendant.vote(last\_particle\_membership)\;
                failed\_progenitors.descendant $\rightarrow$ descendant\;
                dead\_list.add(failed\_progenitors)\;
                
            }
            \BlankLine
            link(descendant, main\_progenitor)\;
        }
        \BlankLine
        \BlankLine
        \For{dead\_progenitor in  dead\_list}{
            grave\_yard.add(dead\_progenitor.voters)\;
        }
        \BlankLine
        \BlankLine
        \BlankLine
        last\_particle\_membership = current\_particle\_membership\;
        progenitor\_list = descendant\_list\;
    }
{\bf end}
\caption{Pseudocode of our merger tree algorithm.}
\end{algorithm}

\bsp
\label{lastpage}
\end{document}